\documentclass[aps,prd,10pt,twocolumn,amsmath,superscriptaddress,amssymb,showpacs,floatfix,nofootinbib,longbibliography]{revtex4-2}

\pdfoutput=1
\usepackage{hyperref}
\usepackage{graphicx}
\usepackage{amsfonts,amsmath,amssymb,bm,bbm}
\usepackage{color}
\usepackage{url}
\usepackage{float}
\usepackage{fontawesome5}
\usepackage{CJK}
\usepackage[export]{adjustbox}
\usepackage{amssymb}
\usepackage{pifont}
\usepackage{tabularx, array, makecell}

\newcommand{\cmark}{\ding{51}} 
\newcommand{\xmark}{\ding{55}} 

\hypersetup{
    pdfnewwindow=true,      
    colorlinks=true,       
    linkcolor=blue,          
    citecolor=blue,        
    filecolor=blue,      
    urlcolor=blue        
}

\newcommand{\orcid}[1]{\begingroup
  \hypersetup{hidelinks}\href{https://orcid.org/#1}{\includegraphics[width=10pt]{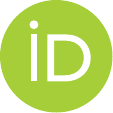}} \endgroup}

\newcommand{\appropto}{\mathrel{\vcenter{
  \offinterlineskip\halign{\hfil$##$\cr
    \propto\cr\noalign{\kern2pt}\sim\cr\noalign{\kern-2pt}}}}}
  
\begin{document}

\title{Angular distribution of gamma rays produced in proton-proton collisions}

\author{Spencer~Griffith \orcid{0009-0002-1988-0768}\,}
\email{griffith.1037@osu.edu}
\affiliation{Center for Cosmology and AstroParticle Physics (CCAPP), Ohio State University, Columbus, Ohio 43210, USA}
\affiliation{Department of Physics, Ohio State University, Columbus, Ohio 43210, USA}

\author{John~F.~Beacom \orcid{0000-0002-0005-2631}\,}
\affiliation{Center for Cosmology and AstroParticle Physics (CCAPP), Ohio State University, Columbus, Ohio 43210, USA}
\affiliation{Department of Physics, Ohio State University, Columbus, Ohio 43210, USA}
\affiliation{Department of Astronomy, Ohio State University, Columbus, Ohio 43210, USA}

\author{Jung-Tsung~Li \begin{CJK*}{UTF8}{bkai}(李融宗)\end{CJK*} \orcid{0000-0003-1671-3171}\,}
\affiliation{Center for Cosmology and AstroParticle Physics (CCAPP), Ohio State University, Columbus, Ohio 43210, USA}
\affiliation{Department of Physics, Ohio State University, Columbus, Ohio 43210, USA}
\affiliation{Department of Astronomy, Ohio State University, Columbus, Ohio 43210, USA}

\author{Annika~H.~G.~Peter \orcid{0000-0002-8040-6785}\,}
\affiliation{Center for Cosmology and AstroParticle Physics (CCAPP), Ohio State University, Columbus, Ohio 43210, USA}
\affiliation{Department of Physics, Ohio State University, Columbus, Ohio 43210, USA}
\affiliation{Department of Astronomy, Ohio State University, Columbus, Ohio 43210, USA}

\date{\today}


\begin{abstract}
Accurate modeling of how high-energy proton-proton collisions produce gamma rays through the decays of pions and other secondaries is needed to correctly interpret astrophysical observations with the Fermi-LAT telescope.  In the existing literature on cosmic-ray collisions with gas, the focus is on the gamma-ray yield spectrum, $d N_\gamma/dE$.  However, in some situations, the joint energy and angular distribution can be observed, so one needs instead $d^2 N_\gamma/dE \, d\Omega$.  We provide calculations of this distribution over the energy range from the pion production threshold to $100~{\rm GeV}$, basing our results on \texttt{FLUKA} simulations.  We provide the results in tabular form and provide a Python tool on GitHub to aid in utilization.  We also provide an approximate analytic formula that illuminates the underlying physics.  We discuss simplified examples where this angular dependence can be observed to illustrate the necessity of taking the joint distribution into account.~\href{https://github.com/skgriffith/Gamma-Ray-Angular-Distribution}{\faGithub}
\end{abstract}

\maketitle


\section{Introduction}
\label{sec:introduction}

Observations of astrophysical gamma-ray fluxes are critical to understanding the origins of cosmic rays (CRs)~\cite{stecker1971cosmic, aharonian2004very, dermer2009high}. Here we focus on hadronic cosmic rays, i.e., protons and light nuclei. Because CRs are charged, and therefore magnetically deflected during propagation, their arrival directions do not reveal their source directions.  Above a total energy of $E_{\rm th} = 1.2~{\rm GeV}$, hadronic CRs can undergo inelastic collisions with ambient gas, producing short-lived secondary particles, including pions.  The subsequent decays of these secondaries produce gamma rays and neutrinos, both of which propagate without deflection.  Many potential CR sources have been identified from gamma-ray observations --- and even had their spectra measured --- though it remains unclear if those sources are accelerating hadronic or leptonic (electron) CRs~\cite{HESS:2005mgd, HUANG2007429, Fermi-LAT:2013iui, HESS:2017jnk, Abeysekara:2021yum, MAGIC:2022rmr}. Ultimately, neutrino observations could break that degeneracy and help reveal the details of CR accelerators~\cite{Sudoh:2022sdk, Sudoh:2023qrz, Song:2023nht, Fang:2024fyd}.

That overall logic can also be reversed.  In situations where we know the CR flux and spectrum, we can use gamma rays to probe the physics of the target.  As one example, we note the surface of the Sun, which is irradiated by hadronic CRs. We have direct measurements of the CR flux and spectrum from observations near Earth~\cite{PhysRevLett.114.171103, PhysRevLett.115.211101, Maurin:2023alp}. At the energies considered, the modulation of cosmic rays due to the heliospheric magnetic field is minimal, so the observations at 1 AU are a reasonable approximation of the spectrum near the Sun~\cite{Li:2022zio}. Due to strong magnetic fields near the solar surface, some fraction of CRs are reflected from ingoing to outgoing before undergoing inelastic collisions in the atmosphere and producing outgoing gamma rays~\cite{Seckel:1991ffa, Orlando:2008uk, Zhou:2016ljf, hudson:2020, Li:2020gch, Gutierrez:2019fna, Gutierrez:2022mor}. A variety of approaches have been considered to model solar gamma-ray observations. As one recent example, Mazziotta et al.~\cite{Mazziotta:2020uey} calculated the yield of gamma rays using the potential field source surface (PFSS) model for the coronal fields, as well as modifications to represent the solar-surface magnetic field. Li et al.~\cite{Li:2020gch} performed similar calculations using PFSS and concluded that additional magnetic structures are necessary to account for the high energy gamma ray emission. Li et al.~\cite{Li:2023twp} and Puzzoni et al.~\cite{Puzzoni:2024enz} focused on the role of smaller magnetic structures near the solar surface. By comparing these and similar calculations to observations, the gamma-ray flux and spectrum observed from the solar disk~\cite{Fermi-LAT:2011nwz, Linden:2018exo, Tang:2018wqp, Linden:2020lvz, Arsioli:2024scu, Linden:2025xom, Acharyya:2025xya} can be used to probe the magnetic structures of the Sun.

In both thrusts above, success depends critically upon an accurate understanding of the particle physics of gamma-ray production.  However, the underlying collisions involve soft hadronic interactions for which we cannot use perturbative quantum chromodynamics. We therefore resort to phenomenological models that describe the data from accelerator experiments and astrophysical observations~\cite{stecker1971cosmic, aharonian2004very, dermer2009high, Badhwar:1977zf, Hanssgen:1982tb, Ranft:1983km, Hanssgen:1983zy, Mohring:1983kk, 1986A&A...157..223D, Hanssgen:1987gw, Capella:1992yb, Kaidalov:2003au, Kamae:2006bf, Karlsson:2007pt}. Many such models exist --- with varying regimes of validity --- and are incorporated, along with interpolations of experimental data, into Monte Carlo simulations to predict the yields of secondaries in nucleon-nucleon interactions.  To simplify the use of these Monte Carlo results, often one uses either tables or approximate functional forms.

In many astrophysical problems, it is enough to consider the observed gamma-ray spectrum, which follows from the yield spectrum for individual collisions, $dN_\gamma/dE$~\cite{Kelner:2006tc, Kafexhiu:2014cua, Orusa:2023bnl}, neglecting the details of the angular distribution.  For source emission, the CR energies are usually assumed to be high enough that relativistic beaming of the secondaries causes the gamma rays to be collinear with the CRs.  And for diffuse emission, e.g., that of the Milky Way plane, where CRs propagate near-isotropically, the angular distribution of the fundamental interaction is unimportant.

In this paper, we consider situations where the angular distribution must be considered as well, i.e., where one needs $d^2 N_\gamma/dE \, d\Omega$.  Work on this topic has been very limited~\cite{Koers:2006dd, Karlsson:2007pt}; in Sec.~\ref{sec:review}, we discuss how we make significant improvements upon it.  The overall goal of our paper is to provide useful phenomenological tools for calculating the angular distribution, as a companion to similar phenomenological tools for the energy distribution~\cite{Kelner:2006tc, Kafexhiu:2014cua, Orusa:2023bnl}.  While we will develop applications in future work, here we note some important cases.

One example of where this is needed is for a CR beam produced in a jet and impacting a target that is viewed off axis.  A second example is the above-mentioned irradiation of the Sun by CRs. The CRs reflected at the solar surface emerge at a variety of angles relative to the observer, so the observed gamma-ray yield will depend on the angular distribution for each interaction.  (A closely related case is irradiation of the Moon by CRs~\cite{Moskalenko:2007mk, Abdo:2012nfa, Fermi-LAT:2016tkg}, where the lack of magnetic fields means that only highly non-collinear gamma rays escape.)  For both examples, because the degree of relativistic beaming depends on energy, collinearity is not a good approximation when the particles are only quasi-relativistic.  

To address these and other needs, we calculate the correlated energy and angular distributions using \texttt{FLUKA}~\cite{Ferrari:2005zk, Bohlen:2014buj, Battistoni:2015epi, Ahdida:2022gjl} simulations to model proton-proton interactions and the decays of secondary particles.  We consider interactions from the kinematic threshold to 100 GeV, a choice motivated by the onset to collinearity.

In Sec.~\ref{sec:review}, we review the particle physics of proton-proton interactions and their phenomenological models.  In Sec.~\ref{sec:Discussion (Non-Analytic)}, we describe our calculations and present our results for the double differential gamma-ray yield per interaction.  \textit{The specific utility of these results is that for complex astrophysical simulations one could avoid using \texttt{FLUKA} directly, greatly improving code flexibility and speed.}  In Sec.~\ref{sec:discussion (analytic)}, we provide an approximate fitting formula.  In Sec.~\ref{sec:observation}, we provide examples of where our results will be important.  Finally, in Sec.~\ref{sec:conclusion}, we conclude and discuss avenues for further progress.


\section{Review of proton-proton collisions}
\label{sec:review}

For CR collisions with gas, the most important case is protons (in CRs) on protons (in gas) because these are the most abundant and we can obtain results for heavier nuclei as an enhancement factor~\cite{Kachelriess:2014mga}. For the nuclear cases, Mazziotta et al.~\cite{Mazziotta:2015uba} provide a database of cross sections for a large variety of projectile-target combinations, but do not consider the angular distribution. Additionally, the \texttt{FLUKA} collaboration has published the results of different projectile-target combinations, including some double-differentially~\cite{Ballarini:2024uxz}, but they do not cover gamma rays in the regime of interest here. Neutrons are typically not present except as secondaries, and neutron-proton collisions are similar to proton-proton collisions.

We limit our attention to gamma rays with $E_{\gamma} \geq 100~{\rm MeV}$. This choice is based on the effective low energy limit of the Fermi-LAT telescope~\cite{2009ApJ...697.1071A}, the observations of which are the primary motivation for this work.


\subsection{Basic physical considerations}
\label{subsec:basic physical considerations}

The production of gamma rays in proton-proton collisions occurs primarily through the decays of secondaries. In this subsection, we consider the simplest case where the only secondaries yielding gamma rays are neutral pions, with more complex final states becoming important at increasing energy (which we take into account in our full calculation in Sec.~\ref{sec:Discussion (Non-Analytic)}). For orientation, we review the basic physics relevant to our study. This includes a discussion of the relevant cross sections as well as an approximate result for the typical gamma-ray energy from simple physical arguments. This will provide a useful comparison for our results from simulations.   

\begin{figure}
\includegraphics[width=0.98\columnwidth]{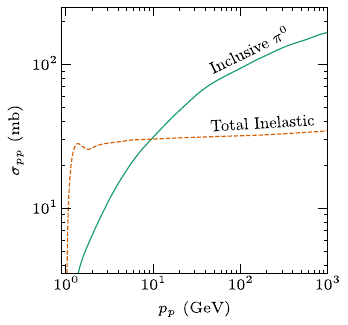}
\caption{Comparison of the (multiplicity weighted) inclusive cross section for $p + p \rightarrow \pi^0 + X$ to the total inelastic proton-proton cross section. The crossover at $p_p \sim 10~{\rm GeV}$ is due to the rising multiplicity in $\pi^0$ production.}
\label{fig:pi0_cross_section}
\end{figure}

Figure~\ref{fig:pi0_cross_section} sets the stage by showing key cross sections for proton-proton collisions~\cite{1986ApJ...307...47D, Workman:2022ynf}.  We are ultimately interested in the gamma-ray yield per inelastic proton-proton collision. Because gamma rays are produced in the decay of neutral pions, the relevant cross sections are the \textit{multiplicity-weighted} inclusive $\pi^0$ production cross section ($p + p \rightarrow \pi^0 + {\rm X}$, where X is any combination of particles) and the total inelastic cross section ($p + p \rightarrow {\rm X}$, where X is any combination of particles, possibly including pions, except strictly $pp$). We focus on the multiplicity-weighted cross section because to obtain the yield we must know the average number of pions produced per inelastic interaction. Below a laboratory momentum of $p_p \sim 10~{\rm GeV}$, the inclusive $\pi^0$ cross section is subdominant.  This reflects the fact that at low energies $\pi^+$ production is dominant, which follows from considerations of the isospin, parity, and angular momentum of the system~\cite{lock1970intermediate}.  Above $\sim$$10~{\rm GeV}$, the $\pi^0$ cross section exceeds the total inelastic cross section due to the rising final-state multiplicity of neutral pions. For reference, temporarily working in the CM frame, the total pion multiplicity (including charged pions) increases with energy, being $\langle N_{\pi} \rangle \sim 5, 10, 15$ at $E_{\rm CM} = 10, 50, 100~{\rm GeV}$~\cite{ParticleDataGroup:2012pjm}.

It is useful to make a rough approximation~\cite{Gaisser:2016uoy} of the average gamma-ray energy resulting from a proton-proton collision, assuming production of a single $\pi^0$. This is generally not the case, except very close the threshold. We discuss the role of the secondary multiplicity below. For clarity, we temporarily restore factors of $\hbar$ (but not $c$) before reverting back to natural units. For an individual parton of the CR as viewed in the lab frame, we have $\Delta x \Delta p \approx \hbar$, where $\Delta x$ is the Lorentz-contracted longitudinal radius of the energetic proton and $\Delta p \approx p_{\rm part}$, where $p_{\rm part}$ is the parton's momentum. Ignoring the transverse component of the parton's momentum,  $p_{\rm part} \approx \hbar/\Delta x$.  Now, $\Delta x$ is related to the uncontracted proton radius ($R_p$) through $\Delta x = R_p / \gamma$, where $\gamma$ is the Lorentz factor, and the total proton energy in the lab frame ($E_p$) is given by $E_p = \gamma m_p$. Therefore, $p_{\rm part} \approx \hbar E_p/ m_p R_p$. If we assume that the momentum transfer during the collision is small and ignore the pion mass, then the average pion energy is the same as the parton energy:
\begin{equation}
    \langle E_{\pi} \rangle \sim \frac{E_p}{5},
    \label{eq:E_pi_av}
\end{equation}
as $\hbar / m_p R_p \approx 1/5$.

These pions decay as $\pi^0 \rightarrow \gamma + \gamma$ with a branching ratio of $\sim 99 \%$~\cite{ParticleDataGroup:2024cfk}.  In the pion's rest frame, there are two back-to-back gamma rays, each with energy $E_{\gamma,\rm {CM}} = m_{\pi} / 2$. In the lab frame, this corresponds to a range of possible energies~\cite{stecker1971cosmic},
\begin{equation}
    \frac{m_{\pi}}{2} \gamma (1 - \beta) \leq E_\gamma \leq \frac{m_{\pi}}{2} \gamma (1 + \beta).
    \label{eq:E_gamma_pi}
\end{equation}
Because the decay occurs isotropically in the rest frame, the distribution between these two limits is flat in $\log{E_\gamma}$ and is centered on
\begin{equation}
    \langle {E_\gamma} \rangle = \frac{E_{\pi}}{2}.
    \label{eq:E_gamma_pi_av}
\end{equation}
If a gamma ray is emitted backwards relative to the pion's direction of travel, then it will be boosted down to $E_\gamma < m_\pi/2 < 100~{\rm MeV}$, below the range we consider.  Therefore, the gamma rays of interest must be forward-going relative to the parent pion. For relativistic pions, these gamma rays are beamed into a cone of size $\Delta \theta \approx 1 / \gamma$. This is the origin of the collinearity mentioned above. However, because the size of the cone scales inversely with energy, near the kinematic threshold the angular distribution is quite broad.

We combine Eqs.~(\ref{eq:E_pi_av}) and~(\ref{eq:E_gamma_pi_av}) to find the average gamma-ray energy from a proton-proton collision:
\begin{equation}
    \langle {E_{\gamma}} \rangle \sim \frac{E_p}{10}.
    \label{eq:E_gamma_av}
\end{equation}
This heuristic approach, which is called the $\delta$-function approximation, is reasonably accurate when considering broad CR and thus broad gamma-ray spectra, as assessed with Monte Carlo simulations (Sec.~\ref{sec:Discussion (Non-Analytic)})~\cite{1996A&A...309..917A, Kelner:2006tc}. One consequence is that a power-law spectrum of CRs induces a power-law spectrum of pions, which in turn induces a power-law spectrum of gamma rays~\cite{stecker1971cosmic}.

However, this simple approach neglects numerous complications, which range from somewhat ignorable to serious.  One example is the increase in pion multiplicity previously mentioned. This complication is somewhat alleviated because there is typically a leading pion that takes the majority of the energy~\cite{Basile:1981ni}. A second, closely related, issue is the increasing relevance of other mesons with increasing energy.  As a third example, near threshold this approximation simultaneously assumes that the pion multiplicity is one (appropriate for low proton energies) and that the pion mass can be neglected (appropriate for high proton energies). For a fourth example, the spread of production and decay energies means that the width of the gamma-ray distribution for a fixed proton energy actually spans a significant energy range, although this depends on the value of $E_p$ and our choice to consider only $E_\gamma \ge 100$~MeV.  For example, at $E_p = 1.33~{\rm GeV}$, which is the lowest energy considered here (just above the pion production threshold at $1.2~{\rm GeV}$), we have $100~{\rm MeV} \leq E_\gamma \leq 266~{\rm MeV}$, while at higher energy ($E_p = 10~{\rm GeV}$) we have $100~{\rm MeV} \leq E_\gamma \leq 9~{\rm GeV}$. For these and other reasons, a more sophisticated treatment of the collision process is needed, especially for the angular distribution, where low-energy, non-collinear emission is expected to be important.


\subsection{Phenomenological modeling with simulations}
\label{subsec:simulations}

To model proton-proton interactions, there are a wide variety of simulation codes, applicable in different energy regimes.  As noted above, we use \texttt{FLUKA}~\cite{Ferrari:2005zk, Bohlen:2014buj, Battistoni:2015epi, Ahdida:2022gjl}.  Before explaining why we make this choice, we briefly review other codes for context, as they are likely more familiar.

\begin{table*}
\begin{tabularx}{\textwidth}{X|X|X|X|X|X|}  
 \cline{2-6}
 & Kelner et al.~\cite{Kelner:2006tc} 
 & Kafexhiu et al.~\cite{Kafexhiu:2014cua} 
 & Karlsson et al.~\cite{Karlsson:2007pt} 
 & Koers et al.~\cite{Koers:2006dd} 
 & This Paper \\
 \hline
 \multicolumn{1}{|X|}{Proton K.E. Range}
 & 100--$10^6$ GeV 
 & 0.28--$10^6$ GeV 
 & 0.49--5$\times 10^5$ GeV 
 & $10^3$--$10^6$ GeV 
 & 0.39--100 GeV \\ 
 \hline
 \multicolumn{1}{|X|}{Detailed Energy Spectra} 
 & \makecell{\raisebox{1.5ex}\cmark} 
 & \makecell{\raisebox{1.5ex}\cmark} 
 & \makecell{\raisebox{1.5ex}\xmark} 
 & \makecell{\raisebox{1.5ex}\xmark} 
 & \makecell{\raisebox{1.5ex}\xmark} \\
 \hline
 \multicolumn{1}{|X|}{Angular Distributions}
 & \makecell{\raisebox{2ex}\xmark} 
 & \makecell{\raisebox{2ex}\xmark} 
 & \makecell{\raisebox{2ex}\cmark} 
 & \makecell{\raisebox{2ex}\cmark} 
 & \makecell{\raisebox{2ex}\cmark} \\
 \hline
 \multicolumn{1}{|X|}{Fundamental Theory} 
 & \makecell{\raisebox{1.5ex}\xmark} 
 & \makecell{\raisebox{1.5ex}\xmark} 
 & \makecell{\raisebox{1.5ex}\xmark} 
 & \makecell{\raisebox{1.5ex}\xmark} 
 & \makecell{\raisebox{1.5ex}\xmark} \\
 \hline
 \multicolumn{1}{|X|}{Hadronic Model(s)} 
 & \makecell{SIBYLL} 
 & \makecell{GEANT4,\\ PYTHIA\:8,\\ SIBYLL\:2.1,\\ QGSJET-I,\\ Data} 
 & \makecell{Kamae et al.\\(2005, 2006)~\cite{Kamae:2006bf}} 
 & \makecell{PYTHIA\:6.3} 
 & \makecell{FLUKA} \\
 \hline
 \multicolumn{1}{|X|}{Analytic Formulas} 
 & \makecell{\cmark} 
 & \makecell{\cmark} 
 & \makecell{\cmark} 
 & \makecell{\cmark} 
 & \makecell{\cmark} \\
 \hline
 \multicolumn{1}{|X|}{Quantified Errors} 
 & \makecell{\cmark} 
 & \makecell{\cmark} 
 & \makecell{\xmark} 
 & \makecell{\cmark} 
 & \makecell{\cmark} \\
 \hline
 \multicolumn{1}{|X|}{Tabular Data Sharing} 
 & \makecell{\raisebox{1.5ex}\xmark} 
 & \makecell{\raisebox{1.5ex}\xmark} 
 & \makecell{\raisebox{1.5ex}\xmark} 
 & \makecell{\raisebox{1.5ex}\xmark} 
 & \makecell{\raisebox{1.5ex}\cmark} \\ [1ex] 
 \hline
\end{tabularx}
\bigskip
\caption{Comparison of this work to key prior papers. Quantified errors indicates some discussion about the accuracy of the analytic formulas, including maximum errors and distribution. An \xmark ~in fundamental theory indicates that the work is based on phenomenological models of data.}
\label{table:comparison}
\end{table*}

At higher energies, common examples of codes used for hadronic collisions include \texttt{PYTHIA}~\cite{bierlich2022}, \texttt{SYBILL}~\cite{Ahn_2009}, and \texttt{QGSJET}~\cite{Ostapchenko:2004ss}, the latter two of which are concerned specifically with cosmic rays. These codes generally work in the limit where the fundamental interactions occur between partons. Consequently, they typically have a high minimum energy and are applicable to the LHC or other similarly energetic scenarios. \texttt{PYTHIA}, for example, is advertised to work above a lab energy of $\sim$$50~{\rm GeV}$; similar restrictions are expected for the other codes.  Kelner et al.~\cite{Kelner:2006tc} used \texttt{SYBILL} and \texttt{QGSJET} to predict the secondary hadron spectra following proton-proton collisions induced by CRs impinging on gas.  Then they predicted the distributions of the resulting decay products --- gamma-rays, electrons, and neutrinos --- through standard kinematics calculations.  They focused on energies above $100~{\rm GeV}$, defaulting to a modified $\delta$-function approximation at lower energies.

At lower energies, where the fundamental interactions are between nucleons, the underlying  physics is less well-characterized theoretically, but there is abundant data to construct phenomenological models~\cite{dermer2009high}.  At low primary energies, $E_{\rm th} \leq E_p \lesssim 3~{\rm GeV}$, $\pi^0$ production occurs primarily through the resonant production of a $\Delta$ baryon, as described by the isobar model. For $E_p \gtrsim 5~{\rm GeV}$, the scaling model provides an accurate description of the cross section. By using a linear combination of the two models in the transition region, $3~{\rm GeV} \lesssim E_p \lesssim 5~{\rm GeV}$, one obtains a reasonable overall description of the pion yield.  Kafexhiu et al.~\cite{Kafexhiu:2014cua} extended the work of Kelner et al.~\cite{Kelner:2006tc} to lower energies, down to the kinematic threshold, by using a combination of accelerator data at low energy and Monte Carlo simulations with \texttt{GEANT4}~\cite{AGOSTINELLI2003250}, a particle propagation code that works at lower energy than the previously mentioned codes.  Importantly, like Kelner et al.~\cite{Kelner:2006tc}, this work did not describe the angular distribution of the resulting gamma-rays.

For our calculations, we use \texttt{FLUKA}~\cite{Ferrari:2005zk, Bohlen:2014buj, Battistoni:2015epi, Ahdida:2022gjl}, another particle propagation code in wide usage, which incorporates a variety of phenomenological and empirical models which are well-calibrated to experimental data.  Other options include \texttt{GHEISHA}~\cite{Fesefeldt:162911} (which is incorporated into \texttt{GEANT4}) and \texttt{UrQMD}\cite{Bleicher:1999xi}, both of which simulate hadronic interactions in the relevant energy regime. \texttt{FLUKA} has several advantages. First, it has been shown to provide accurate predictions for hadronic interactions at low energies (e.g., Ref.~\cite{Heck:2004rq}). Second, it can be set up to directly simulate proton-proton collisions. Third, like \texttt{GEANT4}, it handles the decays of pions and other secondaries. 

As mentioned in Sec.~\ref{sec:introduction}, Refs.~\cite{Koers:2006dd, Karlsson:2007pt} considered the angular dependence of gamma rays produced in proton-proton collisions. We improve upon these results in several ways. First, while Ref.~\cite{Koers:2006dd} is rich in insights, it focused on energies well above the range of Fermi-LAT (and thus the underlying physics is quite different).  Second, Ref.~\cite{Karlsson:2007pt} based its results on the empirical model of Refs.~\cite{Kamae:2004xx, Kamae:2006bf}, whereas we based ours on \texttt{FLUKA}, which has been validated by decades of experimental data.  Third, we provide an extensive discussion of the physical features of our results. Fourth, we make our results readily available in a compiled database of tables with an accompanying Python utility to facilitate incorporation into astrophysical simulations. Finally, we provide a quantification of the error in our analytic formula as well as reasons for the inability of the fit to cover the entire proton energy range and potential ways forward.

Table~\ref{table:comparison} provides comparisons of this work to the most relevant related papers.


\section{Main results: calculation and validation}
\label{sec:Discussion (Non-Analytic)}

\begin{figure*}[t]
\includegraphics[width=0.98\textwidth]{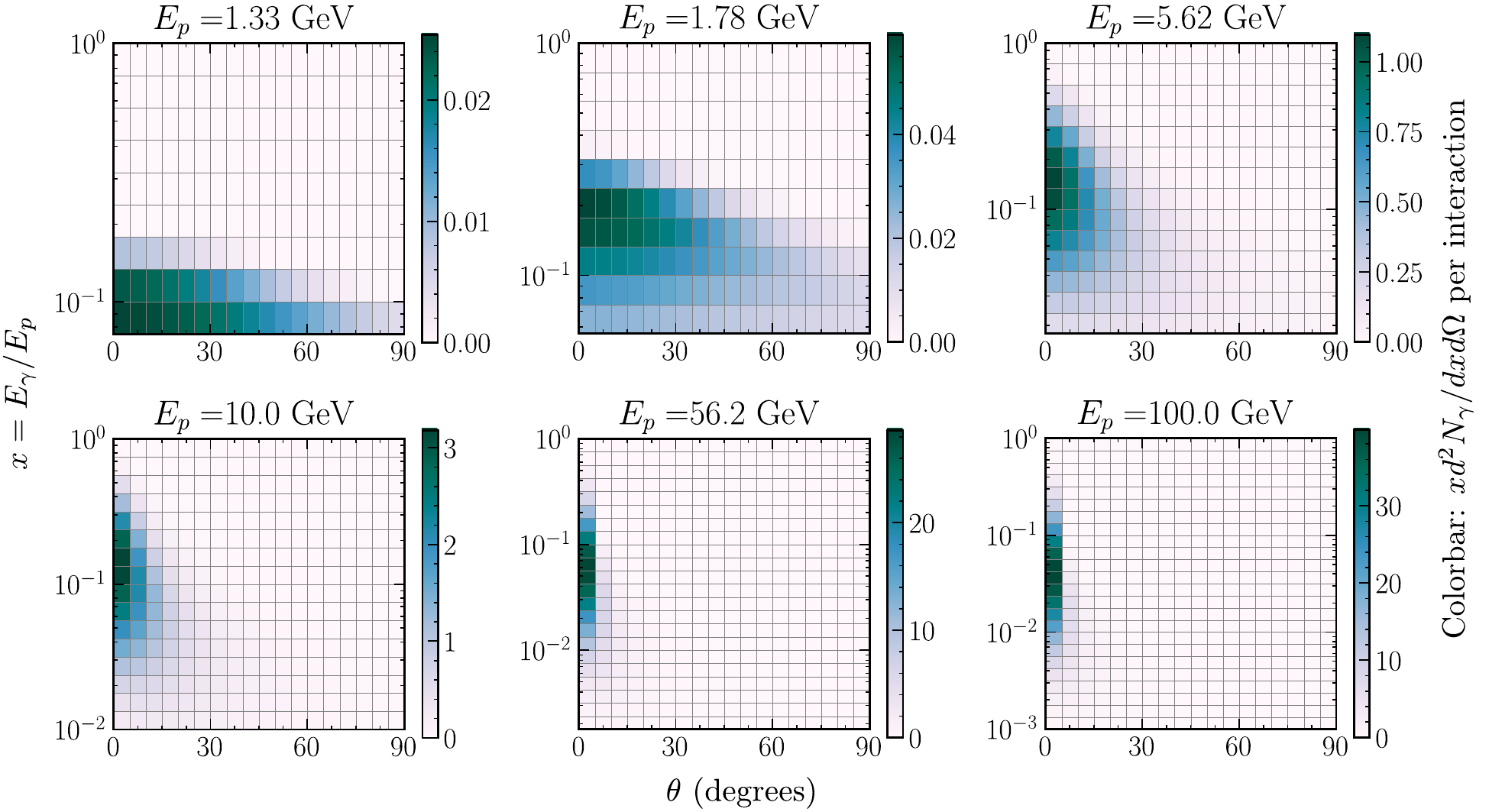}
\caption{Double-differential gamma-ray yields for a representative selection of primary energies, $E_p$.  The approach to collinearity can be seen as $E_p$ increases.  Note that the $x$ ranges and colorbar ranges vary among the panels.}
\label{fig:dd_data_comp}
\end{figure*}

We begin by relating the gamma-ray yield to pion production to connect to the existing literature, primarily Kelner et al.~\cite{Kelner:2006tc} and Kafexhiu et al.~\cite{Kafexhiu:2014cua}. The reader unconcerned with this can skip ahead to Eq.~(\ref{eq:gamma_emissivity}), which describes the gamma-ray emissivity of a monoenergetic proton beam on a point proton target.

For a monoenergetic beam of protons incident on a proton point target, we define $F_{p, \pi}(E_p, E_{\pi}, \theta)$ as the $\pi^0$ yield per steradian at an angle $\theta$ ($0^\circ \leq \theta \leq 180^\circ$) relative to the beam for a single interaction, i.e.,
\begin{equation}
    F_{p, \pi}(E_p, E_{\pi}, \theta) = \frac{d^2 N_{\pi}}{dE_{\pi} d\Omega}.
    \label{eq:F_pi_def}
\end{equation}
If the beam has a number density $n_p$, then there will be $N_{\rm int}$ interactions, where
\begin{equation}
\begin{split}
    N_{\rm int} &= \sigma_{pp} n_p v_p dt, \\
    {\rm with} \,\, v_p &= \frac{\sqrt{E_p^2 - m^2}}{E_p},
    \label{eq:N_int_def}
\end{split}
\end{equation}
where $v_p(E_p)$ is the incident proton velocity and $\sigma_{pp}(E_p)$ is the total inelastic cross section. This gives a total $\pi^0$ emissivity at a given angle of
\begin{equation}
    \frac{d^3 N_{\pi}}{dE_{\pi} dt d\Omega} = \sigma_{pp} F_{p, \pi} v_p n_p.
    \label{eq:pi_emissivity_diff}
\end{equation}
If we allow for a spectrum of incident protons, $dn_p / dE_p$, then the emissivity becomes
\begin{equation}
    \frac{d^3 N_{\pi}}{dE_{\pi} dt d\Omega} = \int_{E_{\rm th}}^{\infty} \sigma_{pp} F_{p, \pi} v_p \frac{dn_p}{dE_p} dE_p.
    \label{eq:pi_emissivity}
\end{equation}
All quantities here are assumed known except for $F_{p, \pi}(E_p, E_{\pi}, \theta)$, which is the quantity of primary interest.

We must now translate this into the gamma-ray emissivity. As discussed in Sec.~\ref{sec:review}, this can be accomplished by considering the kinematics of the pions and computing their decays. However, because \texttt{FLUKA} does this internally, we simply use a modified version of Eq.~\eqref{eq:pi_emissivity}:
\begin{equation}
    \frac{d^3 N_{\gamma}}{dE_{\gamma} dt d\Omega} = \int_{E_{\rm th}}^{\infty} \sigma_{pp} F_{p, \gamma} v_p \frac{dn_p}{dE_p} dE_p,
    \label{eq:gamma_emissivity}
\end{equation}
where $F_{p, \gamma}(E_p, E_{\gamma}, \theta)$, the gamma-ray yield for a single proton-proton interaction, defined similarly to $F_{p, \gamma}(E_p, E_{\pi}, \theta)$ in Eq.~\eqref{eq:F_pi_def}. This has the added advantage of including the gamma rays from other secondaries produced in the collision (kaons, etc.), without needing to treat each species individually.

As the gamma-ray spectrum for a given $E_p$ spans orders of magnitude, it is convenient to work in bins of constant $\Delta \log{E_\gamma}$ (base 10). Because
\begin{equation}
    F_{p, \gamma} \, dE_\gamma = 2.3 \, E_\gamma F_{p, \gamma} \, d \log{E_\gamma},
    \label{eq:F_gamma_to_F_gamma E_gamma}
\end{equation}
we focus on $E_\gamma F_{p, \gamma}$ (noting $\ln 10 = 2.3$). It is also convenient to work in terms of $x = E_\gamma / E_p$, due to the approximate Feynman scaling, which says that $dN/dx$ is a function of $x$ alone (and not also $E_p$) at asymptotically high energy~\cite{Feynman:1969ej}. Therefore, we generally express our results as:
\begin{equation}
    E_\gamma F_{p, \gamma} = x \frac{d^2N_{\gamma}}{dx d\Omega},
    \label{eq:F_gamma_E_gamma_def}
\end{equation}
except where explicitly noted.  This quantity reflects the \textit{number} of gamma rays per log energy bin.  While Fermi-LAT results are often shown as $E^2_\gamma F_{p, \gamma}$ --- the \textit{energy carried} by those gamma rays per log energy bin --- what Fermi-LAT actually measures is $E_\gamma F_{p, \gamma}$.

\begin{figure}
\includegraphics[width=0.98\columnwidth]{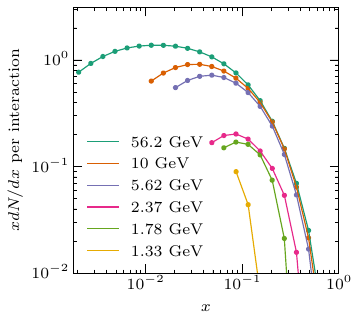}
\caption{Gamma ray yields integrated over solid angle to give the yield spectra for different primary energies, $E_p$.}
\label{fig:ang_proj}
\end{figure}

We performed \texttt{FLUKA} simulations of proton-proton collisions for energetic protons on a static proton target for a range of energies from $E_p = 1.33~{\rm GeV}$ (just above threshold), to $E_p = 100~{\rm GeV}$. The upper limit is based on the increasing collinearity of secondaries due to relativistic beaming, which we discuss below. For the gamma rays, we use eight energies per decade ($\Delta \log{E_\gamma} = 0.125$) and angular bins of $\Delta \theta = 5^\circ$. For convenience, we use the same energy spacing for $E_p$. 

Figure~\ref{fig:dd_data_comp} shows the double-differential results for a representative selection of proton energies. Here we show the angular range limited to $90^\circ$ for clarity. In the following, we first consider the $x$ dependence in the double-differential plots, then integrate over the solid angle to connect to results in the literature. Then we return to the double-differential plots and discuss the angular dependence. Last, we show additional plots that make aspects of the angular behavior more explicit.

In Fig.~\ref{fig:dd_data_comp}, we observe that the most significant gamma ray production occurs at some $x_{\rm peak}$. This is expected from our discussion of the $\delta$-function approximation, where we assigned a specific $x$ value to all gamma rays for a given $E_p$. For $E_p = 1.33~{\rm GeV}$, $x_{\rm peak} \approx 0.08$. Above threshold, $x_{\rm peak}$ rises quickly, and by $E_p = 1.78~{\rm GeV}$, $x_{\rm peak} \approx 0.2$.  This then decreases to $x_{\rm peak} \approx 0.1$ and $x_{\rm peak} \approx 0.06$ for $10~{\rm GeV}$ and $56.2~{\rm GeV}$ protons, respectively (at higher energies, we recover the familiar asymptotic behavior, due to Feynman scaling, of $x_{\rm peak} \approx 0.05$). Ignoring the behavior close to threshold, we see that $x_{\rm peak}$ is a monotonically decreasing function of $E_p$. The value $x_{\rm peak}=0.1$ noted in Sec.~\ref{sec:review} is only appropriate for $E_p$ near 10 GeV. This is due to the simplifying assumptions made in that derivation, mainly that the pion multiplicity is always one (only valid at low energy) and that the pion is massless (only valid at high energy). $E_p \approx 10~{\rm GeV}$ happens to be where both assumptions are sufficiently valid to produce the correct result. The $\delta$-function approximation can work at other energies, but the value of $x_{\rm peak}$ is a function of $E_p$.

Figure~\ref{fig:ang_proj} shows the results for Eq.~\eqref{eq:F_gamma_E_gamma_def} integrated over solid angle. At the highest energy shown, $E_p = 56.2~{\rm GeV}$, the shape is close to the expected scaling relation that occurs for large $E_p$, in which limit the distributions are the same. At the lower energies that we consider, the results fall further below this scaling relation because the interactions are no longer in the deep-inelastic regime.  Importantly, note that the $x_{\rm peak}$ values in our plots are defined for $x dN/dx$, whereas the existing literature frequently focuses on $x^2 dN/dx$.  Our results for each curve abruptly end at low energies because we require $E_\gamma > 100$~MeV, as appropriate for Fermi-LAT.

\begin{figure}
\includegraphics[width=0.98\columnwidth, right]{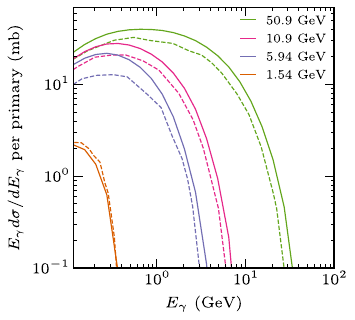}
\caption{Comparison of our results integrated over solid angle (solid lines) to those of Kafexhiu et al. (dashed lines)~\cite{Kafexhiu:2014cua} for a representative selection of proton energies.}
\label{fig:kaf_comp}
\end{figure}

To validate our results, we compare our yield integrated over the solid angle to those described in the existing literature. We focus on the work of Kafexhiu et al.~\cite{Kafexhiu:2014cua} because they also produced results starting from the pion production threshold (though going to higher energies than we do). Importantly, they used a combination of data and other Monte Carlo codes besides \texttt{FLUKA}, so this provides a good test of systematic uncertainties. Furthermore, Kafexhiu et al. continued their analysis to higher energies, where they agreed with the results of Kelner et al.~\cite{Kelner:2006tc}. We also checked (not shown here) that our results agreed with those of Kelner et al.

Figure~\ref{fig:kaf_comp} shows our results integrated over $d \Omega$ and converted to a cross section, i.e., $d \sigma / d E_\gamma$ (which we weight with $E_\gamma$ because of the logarithmic x-axis). Our results for $d N / d E_\gamma$ were obtained by simulating exclusively inelastic collisions. Therefore, the gamma-ray production cross section is obtained by multiplying $d N/ d E_\gamma$ by $\sigma_{\rm inel}$ from Fig.~\ref{fig:pi0_cross_section}. We compare our results (solid lines) to those of Kafexhiu et al. (dashed lines) at a representative selection of energies, finding good agreement, meaning within expected hadronic uncertainties~\cite{Jiang:2020nph, HARP-CDP:2010aip, Uzhinsky:2011qb}. At the lowest proton energy, where both \texttt{FLUKA} and \texttt{GEANT4} are calibrated to extensive experimental data, the agreement is excellent. At $E_p = 5.94$~GeV, we find the greatest divergence. However, near 5 GeV both \texttt{GEANT4} and \texttt{FLUKA} transition between their low and high energy models, so some disagreement in this region is expected. For higher proton energies, some divergence is seen but less than that near 5 GeV.  A significant component of these differences is due to a horizontal shift in the gamma-ray energies, which we conjecture is due to slight differences in the prediction for final-state hadronic multiplicities~\cite{MIPP:2014shj} and/or the well-known uncertainties on forward meson production~\cite{LHCf:2015rcj}, which gives the highest-energy gamma rays. Last, upon integrating our results for $d \sigma / d E_\gamma$ (now including gamma rays below $100~{\rm MeV}$), we find excellent agreement with the inclusive $\pi^0$ cross section from Fig.~\ref{fig:pi0_cross_section}.

Returning to Fig.~\ref{fig:dd_data_comp}, we see that at low $E_p$, gamma-ray production is quite non-collinear. For example, for $E_p < 2~{\rm GeV}$, non-negligible contributions to the yield occur even beyond $\theta = 90^\circ$. As $E_p$ increases, the angular distributions become increasingly collinear.

\begin{figure}
\includegraphics[width=0.98\columnwidth]{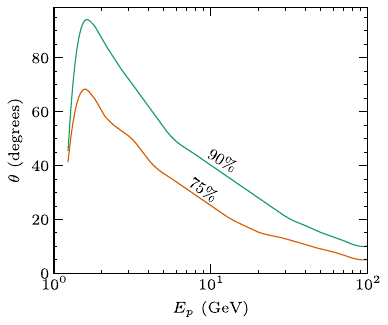}
\caption{Angular cone (defined by the half-angle off the beam axis) necessary to contain $75\%$ (orange) and $90\%$ (green) of the gamma rays above $100~{\rm MeV}$ for proton-proton collisions at different primary proton energies. Note that the curves end abruptly at the threshold energy, $E_{\rm th} = 1.2~{\rm GeV}$.} 
\label{fig:ang_plot}
\end{figure}

Figure~\ref{fig:ang_plot} demonstrates the onset of collinearity more clearly. Here we show the cones, specified by the half angle relative to the beam axis, necessary to contain $75\%$ (orange) and $90\%$ (green) of the gamma rays above $100~{\rm MeV}$.  Already by $E_p \approx 65~{\rm GeV}$, a cone of $10^\circ$ contains $90\%$ of the gamma rays of interest. In most cases, this cone is sufficiently narrow that one can assume that all final-state particles are aligned with the energetic proton, as assumed in the literature.  This justifies using 100 GeV as the maximum of the energy range we consider, since above this energy one can revert to using the collinear analytic formulae of Kelner et al.~\cite{Kelner:2006tc} and Kafexhiu et al.~\cite{Kafexhiu:2014cua}.

\begin{figure*}[t]
\includegraphics[width=0.98\textwidth]{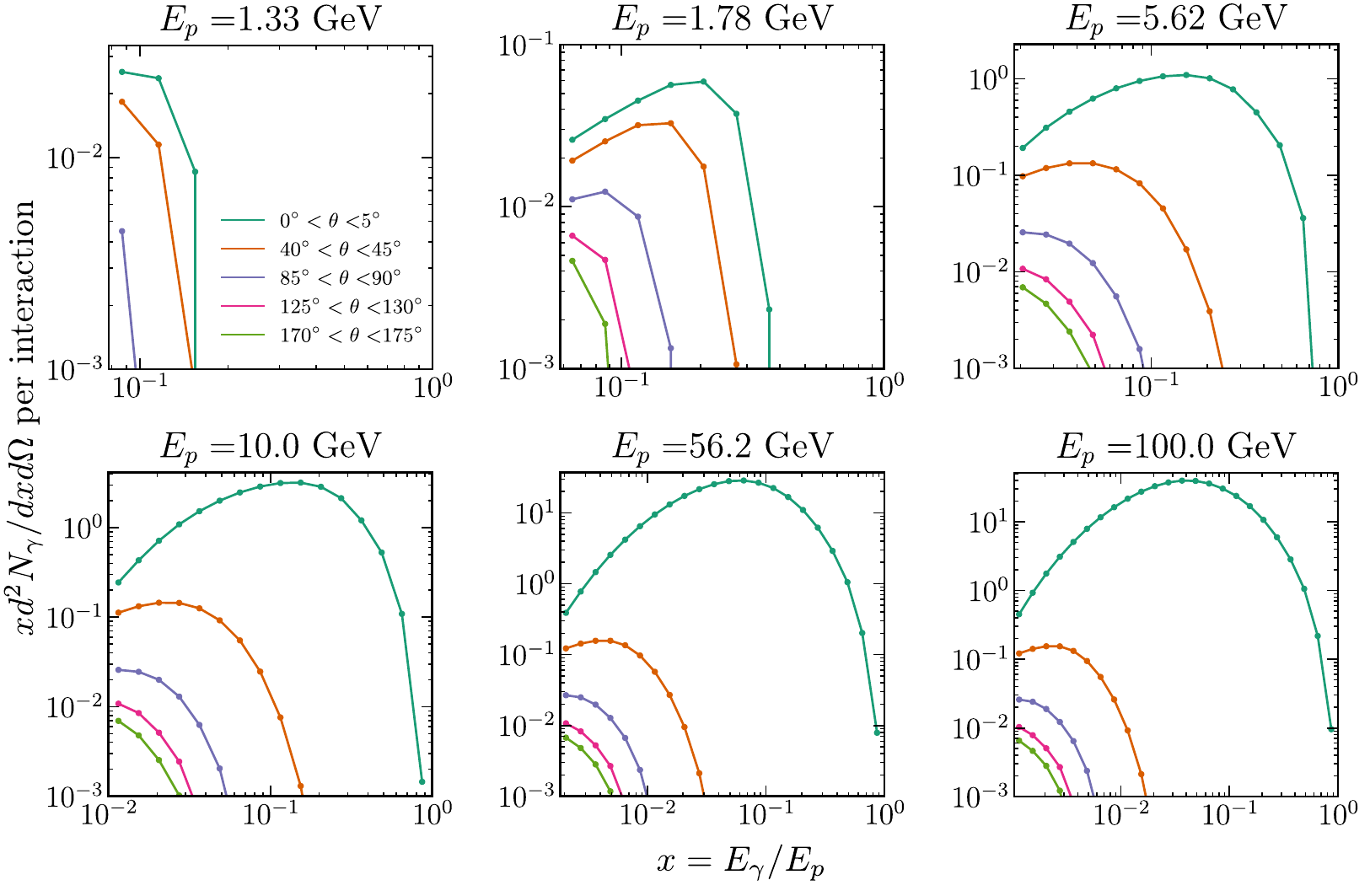}
\caption{Gamma-ray yield above $100~{\rm MeV}$ in different angular ranges relative to the beam direction, $\theta$, for a representative selection of primary energies, $E_p$. For a given $E_p$, we note a distinct spectral shape depending on $\theta$.}
\label{fig:ang_cones}
\end{figure*}

Figure~\ref{fig:ang_cones} shows how the energy distributions of the gamma rays depend on angle.  We show the yield as a function of $x$ in sample angular bins of width $\Delta \theta = 5^\circ$.  The curve with the greatest emissivity is always the most collinear one.  We define ``significant production" in an angular range as having a yield curve within a factor of 10 of the collinear curve at some value of $x$. This is, admittedly, somewhat arbitrary. However, what is actually significant depends on the reader's specific application. We adopt this generic measure to aid in discussion. By $E_p = 56.2~{\rm GeV}$, all significant production occurs within $45^\circ$ of the beam direction, as a result of the overall collinearity of secondaries at high energy. Working downwards in $E_p$, we see the gradual breakdown of collinearity. For $E_p = 5.62~{\rm GeV}$, significant production occurs outside of $45^\circ$, but the emission at angles greater than $90^\circ$ is negligible. At $E_p = 1.78~{\rm GeV}$, for the lowest $x$-bin, production at all angles is within an order of magnitude of the collinear production. For the lowest three $x$-bins, production is significant up to $90^\circ$, while for the lowest five $x$-bins, production is significant up to $45^\circ$. At threshold, for low-$x$ gamma rays, there is significant production at $90^\circ$, and, for all $x$, production is significant at $45^\circ$. 

For a steeply falling proton spectrum (such as for the Galactic cosmic rays, for which it scales as $E_p^{-2.7}$), a large fraction of gamma rays originate from cosmic rays near the kinematic threshold. The analysis above shows that many of the resulting gamma rays will be therefore not be collinear with the parent protons.

Our numerical results are provided as a database of tables and an accompanying Python tool at~\href{https://github.com/skgriffith/Gamma-Ray-Angular-Distribution}{https://github.com/skgriffith/Gamma-Ray-Angular-Distribution}~\cite{Griffith_Angular_distribution_of_2024}. These constitute our primary result and the intended utilization.  When these results are used, one should also cite \texttt{FLUKA}~\cite{Ferrari:2005zk, Bohlen:2014buj, Battistoni:2015epi, Ahdida:2022gjl}.
 

\section{Main results: parameterization}
\label{sec:discussion (analytic)}

There is no established fundamental theoretical framework to predict the details of proton-proton collisions at the energies considered here.  In this section, we fit our numerical results for $x d^2 N/dx d\Omega$ with an ad-hoc functional form. Our intention is that this leads to some insights now and that it may inspire more refined approaches in the future.  We attempt to strike a balance between obtaining reasonable accuracy and limiting the number of parameters. Because we are concerned primarily with departures from collinearity, we choose to optimize the fit over the first decade of $E_p$ above the pion production threshold.  Further, while we start with a more complicated function, we show how it can be simplified to extract intuitive physical insight.

There are two challenges.  First, the yield is a two-dimensional function which varies by orders of magnitude.  Second, for a given $E_p$, a significant portion of the phase space in $x$ and $\theta$ is kinematically forbidden or extremely suppressed.

We address these complications in two ways. We limit our consideration to a reduced $\theta$ range of $0^\circ$ to $90^\circ$, excluding a large portion of the extremely low-yield phase space. This simplification is only reasonable when approximately collinear emission is possible, because the $\theta > 90^\circ$ portion of the phase space becomes irrelevant. In situations where all gamma-ray emission is highly non-collinear (such as lunar gamma rays, mentioned in Sec.~\ref{sec:introduction}), this approximation is not valid. We must also be careful about how we evaluate the quality of the fit. A reasonable measure is the fractional discrepancy between the fit and the data:
\begin{equation*}
    \left| \frac{{\rm data} - {\rm fit}}{\rm data} \right|.
    \label{eq:frac_error}
\end{equation*}
Hereafter we call this quantity the error. In kinematically suppressed bins, both the data and fit will be extremely small. However, the error in these bins may be large due to the division by a small quantity. This is irrelevant to the quality of the fit because these bins contribute negligibly to the yield. We deal with this by designating a cutoff value below which we consider any yield to be insignificant. 

We choose the cutoff value, $n_{\rm cut}$, as the yield such that, for a given $E_p$, if it were present in every bin and integrated over the entire phase space, limited to $0^\circ < \theta < 90^\circ$, we would only over predict a single gamma ray for every 100 interactions:
\begin{equation}
    10^{-2} = \int_{x_{\rm min}}^{1} \int_{0}^{2 \pi} n_{\rm cut} \, d\log x \, d\Omega, \\
    \label{eq:cutoff_int}
\end{equation}
so
\begin{equation}
    n_{\rm cut} = - \frac{10^{-2}}{2 \pi \log{x_{\rm min}}}.
    \label{eq:cutoff}
\end{equation}
We consider any bin where both the data and fit are less than $n_{\rm cut}$ to both be zero (and consequently have zero error). This choice is somewhat arbitrary and may not be appropriate for every application. However, some cutoff is necessary and this is a reasonable one for general purposes. We then measure the error in any bin as:
\begin{equation}
    {\rm error} = \begin{cases}
    0 & {\rm data~and~fit} < n_{\rm cut} \\
    \left| \frac{{\rm data} - {\rm fit}}{\rm max(data,~fit)} \right| & {\rm otherwise}.
    \end{cases}
    \label{eq:error}
\end{equation}
The denominator in the non-zero case is chosen to avoid dividing by zero.

A less obvious complication stems from the change in average pion multiplicity. Near threshold, only one pion can be produced, where this pion has a low momentum. The resulting gamma rays will be nearly isotropically distributed. As $E_p$ increases --- but staying below the threshold for production of two pions --- the resulting pion will have a greater momentum, and thus more forward boosting.  When we reach the threshold to produce two pions, we again obtain low-momentum pions and isotropically distributed gamma rays (whenever two pions are produced instead of one high-momentum pion). This process repeats as the thresholds to produce greater numbers of pions are exceeded.  We find that, for any reasonably parameterized fit, the fit fails for particular values of $E_p$ where these discrete decreases in boosting occur. Once  $E_p$ exceeds approximately 4~GeV, this is no longer a problem because so many multi-pion channels are open.

We present the fit in a modular form to make the important features apparent:
\begin{equation}
    f(x, \theta, E_p ) = A(x, E_p) \exp{\left( -B(x, E_p) \theta^{I(E_p)} \right)},
    \label{eq:fit_eq}
\end{equation}
\begin{equation}
    \left(x \frac{d^2N}{dx d\Omega} \right)_{\rm fit} = 
    \begin{cases}
    f(x, E_p) & {\rm if}~f > 0 \\
    0 & {\rm if}~f \leq 0.
    \end{cases}
    \label{eq:fit_main}
\end{equation}
The piecewise definition is only relevant for $E_p \lesssim 2~{\rm GeV}$, where $f(x, E_p)$ takes on small negative values for bins with large $x$ and small $\theta$. 

The functional form is not theoretically motivated. Rather, it was built by examining individual ``slices" over the phase space and finding distributions to match. This was necessary for two reasons. First, we are not aware of an existing theoretically motivated form for the angular distribution. Existing forms that describe the integrated yield are not readily reverse engineered to a double differential form. Second, fitting the large number of parameters requires some simplicity in the function. The form of this equation was not specifically constructed to address the concern about the discrete pion multiplicity changes. Instead, it is simply a form that was found to be able to match the data reasonably over the energy range of interest.

The functions $A(x, E_p)$, $B(x, E_p)$, and $I(E_p)$ are themselves complicated functions that we present in a modular form:
\begin{equation}
\begin{split}
    &A(x, E_p) = A_0(E_p) e^{-A_1(E_p)x^2} - A_2(E_p) e^{-A_3(E_p)x} \\
    &A_0(E_p) = \frac{A_{00} E_p^2 + A_{01} E_p + A_{02}}{E_p + A_{03}} \\
    &A_1(E_p) = A_{10} e^{-A_{11} (E_p + A_{12})^2} + A_{13} e^{-A_{14} E_p} \\
    &A_2(E_p) = \frac{A_{20} E_p^2 + A_{21} E_p + A_{22}}{E_p + A_{23}} \\
    &A_3(E_p) = A_{30} (E_p + A_{31})^4 e^{-A_{32} E_p} + A_{33} e^{-A_{34} E_p} + A_{35}
    \label{eq:fit_Asub}
\end{split}
\end{equation}
\begin{equation}
\begin{split}
    &B(x, E_p) = B_0(E_p) x + B_1(E_p) \\
    &B_0(E_p) = \frac{B_{00} E_p^2 + B_{01} E_p + B_{02}}{E_p + B_{03}} \\
    &B_1 (E_p) = B_{10} (E_p + B_{11})^{5/7} + B_{12}
    \label{eq:fit_Bsub}
\end{split}
\end{equation}
\begin{equation}
    I(E_p) = \frac{E_p + I_0}{E_p + I_1} + I_2.
    \label{eq:fit_Isub}
\end{equation}
Table~\ref{table:parameters} shows the numerical constants $A_{ij},~B_{kl},~I_m$.

\begin{table}
\centering
\begin{tabular}{||c | c | c | c | c | c||} 
 \hline
 A & Value & B & Value & I & Value \\ [0.5ex] 
 \hline\hline
 $A_{00}$ & 1.69 & $B_{00}$ & 0.0557 & $I_0$ & 0.288 \\ 
 $A_{01}$ & -5.96 & $B_{01}$ & -0.144 & $I_1$ & -0.343 \\
 $A_{02}$ & 6.82 & $B_{02}$ & 0.117 & $I_2$ & 0.182 \\
 $A_{03}$ & 2.71 & $B_{03}$ & -0.545 & & \\
 $A_{10}$ & 42.2 & $B_{10}$ & 0.0013 & & \\
 $A_{11}$ & 2.24 & $B_{11}$ & -4.92 & & \\
 $A_{12}$ & -1.25 & $B_{12}$ & 0.0011 & & \\
 $A_{13}$ & 9.25 & & & & \\
 $A_{14}$ & 0.0280 & & & & \\
 $A_{20}$ & 1.61 & & & & \\
 $A_{21}$ & -6.10 & & & & \\
 $A_{22}$ & 7.38 & & & & \\
 $A_{23}$ & 1.29 & & & & \\
 $A_{30}$ & 0.405 & & & & \\
 $A_{31}$ & -1.87 & & & & \\
 $A_{32}$ & 0.783 & & & & \\
 $A_{33}$ & 113 & & & & \\
 $A_{34}$ & 2.18 & & & & \\
 $A_{35}$ & 5.50 & & & & \\ [1ex] 
 \hline
\end{tabular}
\bigskip
\caption{Parameters for Eqs.~(\ref{eq:fit_Asub}), (\ref{eq:fit_Bsub}), and (\ref{eq:fit_Isub}).}
\label{table:parameters}
\end{table}

\begin{figure*}[t]
\includegraphics[width=0.98\textwidth]{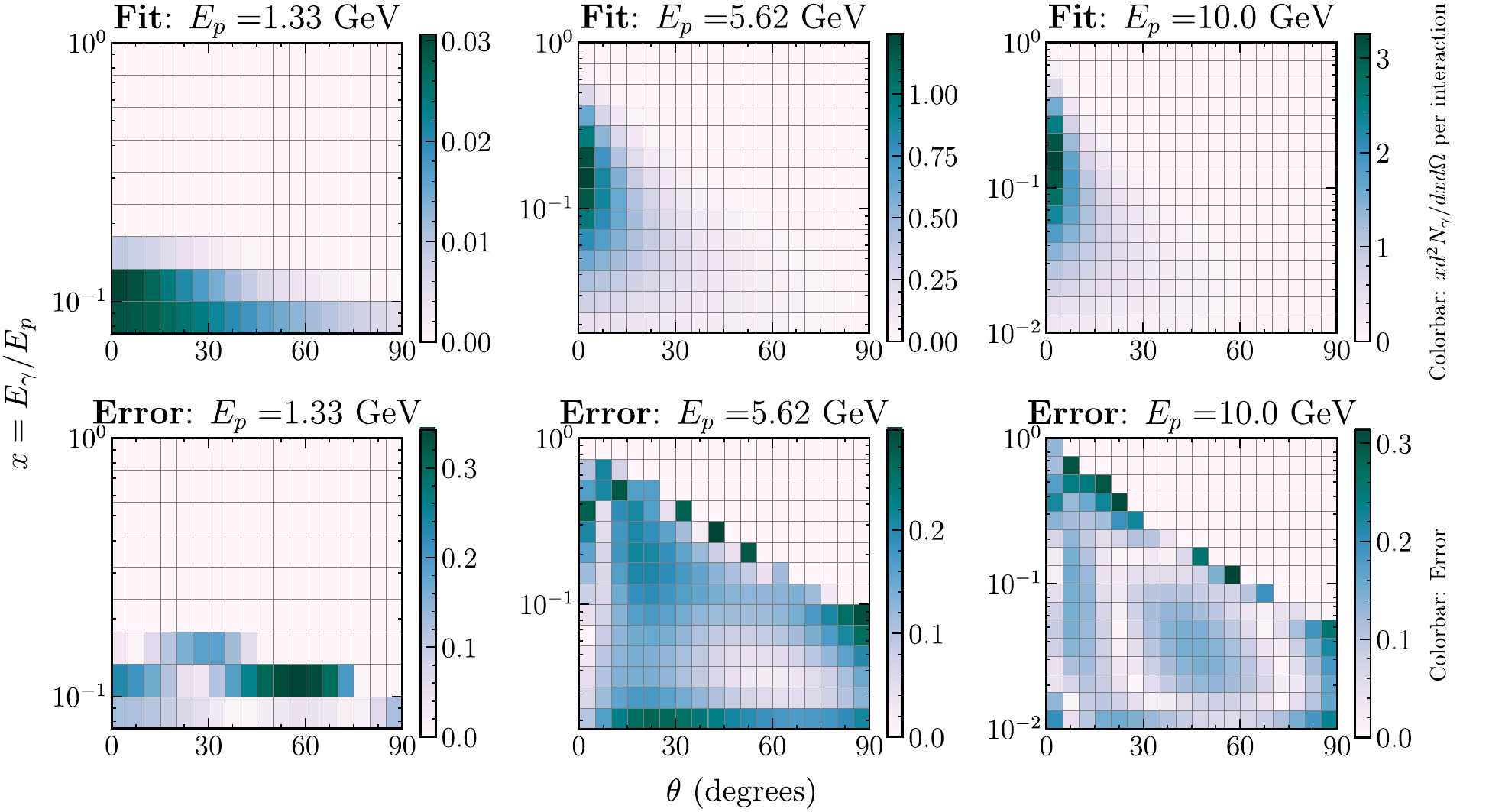}
\caption{Analytic function modeling $x d^2 N / dx d \Omega$, Eq.~(\ref{eq:fit_main}), (top row) and the relative error compared to the \texttt{FLUKA} results, Eq.~(\ref{eq:error}) (bottom row). We note that the largest errors for a given $E_p$ always occur in relatively small-yield bins.}
\label{fig:fit}
\end{figure*}

Figure~\ref{fig:fit} shows the fits and their associated errors for a representative selection of energies. We note that over the entire decade of $1.33~{\rm GeV} \leq E_p \leq 13.3~{\rm GeV}$, the maximum error in the fit is under 40\% (not considering the individual $E_p$ values below $4.22~{\rm GeV}$ where the multiplicity rapidly changes the collinearity and the error increases). The error increases to a maximum of $\sim 65\%$ for $E_p = 17.8~{\rm GeV}$ and $\sim 70\%$ for $E_p = 23.7~{\rm GeV}$. The increase in error is discussed below, but for now we focus on $E_p \leq 13.3~{\rm GeV}$.

\begin{figure}
\includegraphics[width=0.98\columnwidth]{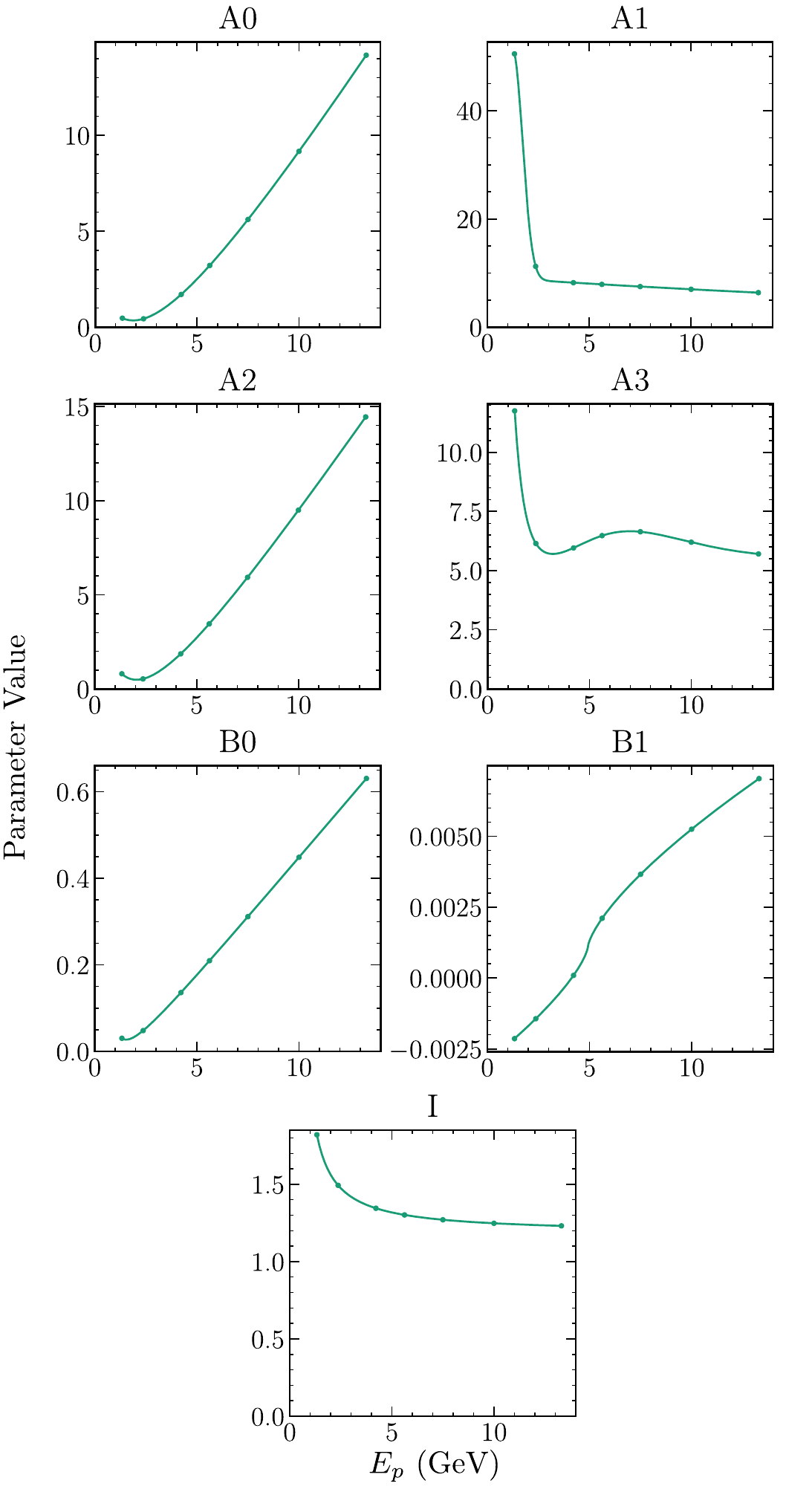}
\caption{Parameters used in Eqs.~(\ref{eq:fit_Asub}), (\ref{eq:fit_Bsub}), and (\ref{eq:fit_Isub}).}
\label{fig:parameters}
\end{figure}

We can gain some insight by examining the values of the parameter functions ($A_0$, $A_1$, etc...) shown in Fig.~\ref{fig:parameters}. We note that only $A_1$ and $A_3$ vary significantly over the decade of $E_p$ under consideration. Furthermore, $A_1$ and $A_3$ vary significantly only for $E_p$ values near threshold. We also note that $A_0 \approx A_2$ and $B_1 \approx 0$. Therefore:
\begin{equation}
    \left(x \frac{d^2N}{dx d\Omega} \right)_{\rm fit} \approx A_0 (e^{-A_1 x^2} - e^{-A_3 x}) \, e^{-B_0 x \theta^I}
    \label{eq:fit_approx}
\end{equation}

This is not to say that these parameters are actually degenerate. Parameterizing the fit in this simplified way significantly increases the associated maximum errors. The extra parameters generally serve to fine-tune the behavior of the function as it approaches zero along the boundary of the non-zero phase space and to accommodate proton energies extremely close to threshold. This is just a useful approximation to discuss the significant features of the overall behavior.

The primary trend, carried by the $e^{-B_0(E_p) x \theta^{I(E_p)}}$ term, is a yield which decreases exponentially with increasing $x$ and $\theta$. Because $I > 1$, we see that the decrease is more rapid with increasing $\theta$ than increasing $x$. The term $(e^{-A_1 x^2} - e^{-A_3 x})$ serves to shift the location of the peak away from $x = 0$ to higher $0 < x < 1$, as we saw in Sec.~\ref{sec:Discussion (Non-Analytic)}.

There are generally a small number of bins with significantly higher relative error. As $E_p$ increases, these bins become relegated to high $x$, low $\theta$ bins or high $\theta$ bins. These bins are always relatively low yield and this trend becomes more pronounced with increasing primary energy. This suggests, and our analysis confirms, that the fit remains accurate significantly past the decade considered here if we choose to limit our range or are willing to accept error in relatively insignificant bins (this is equivalent to increasing the cutoff value in Eq.~\eqref{eq:cutoff}). The details of Fig.~\ref{fig:fit} indicate, and further analysis confirms, that we could add additional parameters and achieve greater accuracy. However, extending the fit also limits its usefulness. Increased parameterization would obscure the physics and limiting the range would render it useless for calculational purposes. For this reason, we chose not to pursue further refinement. However, we encourage further efforts towards describing the yield analytically.


\section{Discussion of applications}
\label{sec:observation}

In this section, we sketch some examples of how the non-collinearity of gamma-ray emission could affect astrophysical observations.  We begin with a jet scenario and then consider solar gamma rays.


\subsection{CRs in astrophysical jets}

If a gas cloud is illuminated from one direction by a CR beam (e.g., produced in an astrophysical jet), then the gamma-ray spectrum we observe depends on our viewing angle relative to the jet axis. Returning to Fig.~\ref{fig:ang_cones}, this figure describes the situation for monoenergetic CR sources where the gas cloud is approximated by a point target, as is appropriate if the column density is not large. 

Compared to on-axis emission, the spectra become noticeably narrower with increasing angle. Importantly, we see that, in principle, \textit{the spectrum shape allows us to determine the angle of the beam}. This is valuable because in most astrophysical situations the location of the CR accelerator is unknown and of primary interest. We also note that, even for large proton energies and large angles, there is always non-zero gamma-ray emission at low energy. In Fig.~\ref{fig:ang_cones}, the low-energy spectra are truncated by our specific choice of a 100-MeV cutoff for Fermi-LAT; without that, we would see the spectra turn over at lower energies for all angles.   

A more realistic situation is that of a gas cloud illuminated by a beam of CRs with an energy spectrum, e.g., a power law where $\phi_p \propto E_p^{-\alpha}$. In this case:
\begin{equation}
    \begin{split}
    E_\gamma \frac{d^2N}{dE_\gamma d\Omega} &\propto \int_{E_{\rm th}}^\infty \left( E_\gamma \frac{d^2N}{dE_{\gamma} d\Omega} \right)_{E_p} E_p^{-\alpha} dE_p \\
    &\appropto \sum_{E_p} \Delta \log{E_p} E_p^{-\alpha + 1} \left( E_\gamma \frac{d^2N}{dE_{\gamma} d\Omega} \right)_{E_p}
    \end{split}
    \label{eq:total_gamma1}
\end{equation}
where the the term $\left( E_\gamma d^2N dE_{\gamma} / d\Omega \right)_{E_p}$ is the yield from a specific CR energy. Here the sum is over a discrete set of proton energies. We choose the energies with a constant $\Delta \log{E_p}$ so that
\begin{equation}
    E_\gamma \frac{d^2N}{dE_\gamma d\Omega} \appropto \sum_{E_p} E_p^{-\alpha + 1} \left( E_\gamma \frac{d^2N}{dE_{\gamma} d\Omega} \right)_{E_p}
    \label{eq:total_gamma2}
\end{equation}

\begin{figure}
\includegraphics[width=0.98\columnwidth]{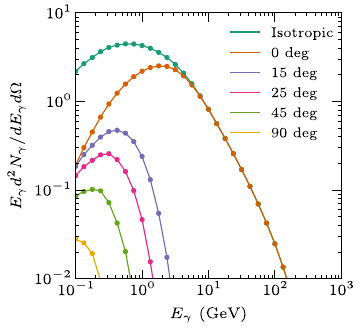}
\caption{Gamma-ray spectrum for a target illuminated by a CR spectrum with power-law index of 2.2 and ranging up to $1000~{\rm GeV}$. We show results for different viewing angles relative to the beam (and also isotropic illumination).}
\label{fig:comb_spec_plot}
\end{figure}

Figure~\ref{fig:comb_spec_plot} shows the right-hand side of Eq.~\eqref{eq:total_gamma2} for a CR spectrum with $\alpha = 2.2$ that extends up to an arbitrary high-energy cutoff ($1000~{\rm GeV}$). The choice of $\alpha$ used here for demonstration is informed by the spectrum produced by Fermi acceleration. At small angles, we again find a broad spectrum and a steep initial rise in yield with increasing $E_\gamma$ above $100~{\rm MeV}$. By $45^\circ$, we see that the spectrum is essentially flat above $100~{\rm MeV}$ before dropping sharply and producing a much narrower spectrum. We also note that for an isotropic source, such as a diffuse CR population impinging on a gas cloud, we see a different spectrum than we observe for any of the individual viewing angles.  As above, in principle the shape of the gamma-ray spectrum could determine the direction of the beam from the source, assuming that we understand the original CR spectrum.  Further, it could determine if the source is beamed or not (see the Isotropic case in Fig.~\ref{fig:comb_spec_plot}).

\subsection{Solar gamma rays}

The production of gamma rays through irradiation of the solar disk is another example of where non-collinear effects matter.  If the Sun's magnetic fields were neglected, then the observed distribution of gamma rays on the solar surface would depend entirely on the interplay of the angular distribution and the geometry of the target. For very high energy gamma rays, emission is collinear and necessarily due to very high energy CRs. In this case, only protons directly pointed at the observer and that graze the solar surface would produce observed gamma rays, e.g., only the limb of the Sun (a thin ring) would be bright in gamma-ray emission. As the gamma-ray energy decreases, we draw increasingly from the non-collinear part of the emission that occurs for all proton energies, so that the thin ring around the Sun would become an annulus.  As the gamma-ray energies decrease even further, we would observe emission over the full disk. However, the emitted spectrum would depend on the radius from the disk center. These predictions for the spatial distribution of the gamma rays differ from what is observed, indicating that CR interactions with solar magnetic fields must be important~\cite{Zhou:2016ljf}.

When we stop neglecting the solar magnetic fields, something very different happens.  In this case, the most relevant CRs are those arriving from the \textit{front side} of the Sun, not the back side as above.  The observed gamma rays are produced by CRs that are first reflected by solar magnetic fields, then interact with gas in the solar atmosphere on their way back out~\cite{Seckel:1991ffa, Zhou:2016ljf, hudson:2020, Mazziotta:2020uey, Li:2020gch, Gutierrez:2019fna, Gutierrez:2022mor, Li:2023twp, Puzzoni:2024enz}.  At very high gamma-ray energies, where the emission is collinear, only outgoing protons pointed directly towards the observer are important.  These could arise from locations across the full solar disk, with the distribution depending on the magnetic field structures.  These very high energy gamma rays will be accompanied by a spectrum that reflects the on-axis components shown in Fig.~\ref{fig:ang_cones}.  The power-law spectrum of the gamma rays would reflect the power-law spectrum of the CRs that make it to the solar surface (a harder spectrum than seen near Earth).  As the gamma-ray energies decrease, non-collinear emission contributes an increasing fraction of the observed gamma rays. These can also arise from locations across the solar disk, from protons that are redirected to be generally outgoing from the solar surface, but not directly towards the observer, before interacting. This effect is more pronounced if mirroring only occurs in certain parts of the disk. As an extreme case, one can consider coronal mass ejections or solar flares, which are both highly localized~\cite{Fermi-LAT:2013vao, Pesce-Rollins:2015hpa, Kouloumvakos:2020ces}. The details depend on modeling of the solar magnetic fields. The effects described here become especially interesting if our ability to resolve where gamma rays originate on the solar disk improves. 


\section{Conclusions}
\label{sec:conclusion}

Observations of the gamma-ray sky provide an important tool to understand the sources of hadronic cosmic rays. These gamma rays follow from the decays of pions produced in the collisions between the CRs and ambient gas. In this paper, we focused exclusively on the proton-proton case because these make up the bulk of CRs and ambient gas. Consideration of heavier nuclei can be accomplished through nuclear enhancement factors. To properly interpret the gamma rays thus produced, one requires knowledge about the gamma-ray distribution from these proton-proton collisions. In many cases, it is either reasonable to assume that the gamma rays are collinear with their parent CRs, or that the angular distribution does not matter due to isotropy. However, this is not true for all sources. We presented two such examples: gamma rays produced in an off-axis CR jet and gamma rays produced in the Sun's atmosphere.  To properly model the physics of such sources, we need $d^2 N_\gamma/dE \, d\Omega$ instead of just $d N_\gamma/dE$.

We calculated the joint energy and angular distribution of gamma rays produced in proton-proton collisions from the pion production threshold to $100~{\rm GeV}$ using the \texttt{FLUKA} Monte Carlo code. Very little work has been done previously on the angular distribution~\cite{Koers:2006dd, Karlsson:2007pt}, and it had the limitations noted in Sec.~\ref{sec:review}.  Before proceeding, we checked that our calculations, when integrated over angles, match the $d N_\gamma/dE$ results of the well-known paper by Kafexhiu et al.~\cite{Kafexhiu:2014cua}, which are based on other Monte Carlo codes as well as experimental data.

Our main results are shown in Fig.~\ref{fig:dd_data_comp}.  At high proton energies (several tens of GeV), most of the emission, especially the highest-energy gamma rays, can be treated as collinear.  The validity of this approximation is quantified in Fig.~\ref{fig:ang_plot}.  However, at lower energies, there is a significant production of gamma rays off axis. These gamma rays are produced with a different spectrum concentrated at low energies. This needs to be taken into account to accurately interpret the physics of sources that are not isotropic.  Further, even for high-energy protons, there is always low-energy gamma-ray emission with a reduced flux and narrower spectrum, as shown in Figs.~\ref{fig:ang_cones} (for the monoenergetic CR case) and~\ref{fig:comb_spec_plot} (for the power-law CR spectrum case). This is due to the kinematics of pion production and decay at low energies.  This effect is related to that used to produce ``off-axis" neutrino beams at terrestrial accelerators~\cite{Para:2001cu}.

Our results are available in tabular form as a database on~\href{https://github.com/skgriffith/Gamma-Ray-Angular-Distribution}{GitHub}~\cite{Griffith_Angular_distribution_of_2024}.  We include a simple Python module which illustrates the usage of the database.  As noted above, any use of our results should also cite \texttt{FLUKA}~\cite{Ferrari:2005zk, Bohlen:2014buj, Battistoni:2015epi, Ahdida:2022gjl}.

We also present an analytic formula that reasonably approximates the gamma-ray yield as a function of angle relative to the beam direction over the first decade of energy above threshold. We discussed the general behavior of this function, as well as the difficulties present in finding an analytic fit valid over the entire energy range. It is our hope that this discussion will spur interest in determining such a function.

Finally, we discussed the observational consequences of the angular dependence. This was accomplished in part through the case of a proton beam incident upon a point target that is viewed off axis. An avenue for continued exploration involves making this model more realistic for astrophysical environments. This includes using extended targets of a given proton density and considering heavier nuclei both in the beam and target. We also qualitatively considered the application of these results to the ongoing question of solar gamma-ray emission, which is also a promising direction for future applications.


\section*{Acknowledgments}

We are grateful for helpful discussions with Florian Herren, Obada Nairat, Chingam Fong, and especially Ervin Kafexhiu. The work of S.~G. was supported by NASA Grant No.\ 80NSSC20K1354 and National Science Foundation Grant No.\ PHY-2310018.  J.~F.~B. was supported by National Science Foundation Grant No.\ PHY-2310018. J.~T.~L. and A.~H.~G.~P. were supported by NASA Grant Nos.\ 80NSSC20K1354 and 80NSSC22K0040.  



\clearpage
\twocolumngrid
\bibliography{bibliography}

\end{document}